\documentclass[amsmath,amssymb,aps,prd,11pt,tightenlines,superscriptaddress,  
nofootinbib,preprintnumbers,notitlepage]{revtex4-1} 
\usepackage{graphicx}  % needed for figures  
\usepackage{dcolumn}   % needed for some tables   
\usepackage{bm}        % for math   
\usepackage{amssymb}   % for math   
\usepackage{amsmath}   
\usepackage{mathrsfs}  %for \mathscr       
\usepackage{graphicx,color} %for color 
\usepackage{dsfont} 
\usepackage[english]{babel}     
\usepackage{graphicx,wrapfig,lipsum}        
\usepackage{multimedia}      
\usepackage[mathscr]{eucal}   
\usepackage{bm}              
\usepackage{amssymb}        
\usepackage{amsmath}   
\usepackage{mathrsfs}       
\definecolor{ao(english)}{rgb}{0.0, 0.5, 0.0}    

\usepackage{sidecap} 
\usepackage{wrapfig}    
\usepackage{subfig} 
\usepackage{float}    
 
\usepackage{ulem}
\begin{document}      
 
 \title{3D Yang-Mills confining properties from a non-Abelian ensemble perspective}

\author{D. R. Junior}
\email{davidjunior@id.uff.br}
\author{L. E. Oxman}  
\email{leoxman@id.uff.br} 
\author{G. M. Sim\~oes}
\email{gustavomoreirasimoes@id.uff.br} 
\affiliation{ 
Instituto de F\'isica, Universidade Federal Fluminense,
Campus da Praia Vermelha, Niter´oi, 24210-340, RJ, Brazil.
}  

\date{\today}

\begin{abstract}     

 In this work, we propose a $3D$ ensemble measure for center-vortex worldlines and chains equipped with 
 non-Abelian degrees of freedom. We derive an effective field description for the center-element average where the vortices get represented by $N$ flavors of effective Higgs fields transforming in the fundamental representation.  
This field content is required to accommodate fusion rules where $N$ vortices can be created out of the vacuum. The inclusion of the chain sector, formed by center-vortex worldlines attached to pointlike defects, leads to a discrete set of $Z(N)$ vacua. This type of SSB pattern supports the formation of a stable domain wall between quarks, thus accommodating not only a linear potential but also the 
L\" uscher term. Moreover, after a detailed analysis of the associated field equations, the asymptotic string tension turns out to scale with the quadratic Casimir of the antisymmetric quark representation. 
These behaviors reproduce those derived from Monte Carlo simulations in $SU(N)$ $3D$ Yang-Mills theory, which lacked understanding in the framework of confinement as due to percolating magnetic defects. 

\end{abstract}

\maketitle
\section{Introduction}

The Wilson loop average in pure SU(N) Yang-Mills theory computed via Monte Carlo simulations displays many important features. Among them, the confinement of fundamental quarks at large distances is signaled by an area law \cite{Bali}.   At intermediate distances, within the confining regime, the simulations show Casimir scaling \cite{Bali2}. That is,   the ratios of the string tensions for quarks in a group representation ${\rm D}$ and the fundamental is given by the ratio of the corresponding quadratic Casimir operators, namely,
\begin{equation}
\sigma_D/\sigma_F = {\rm Tr} \big( {\rm D}(T_A )  {\rm D}(T_A) \big)/{\rm Tr} \big( T_A T_A \big),
\end{equation}
where $T_A $, $A=1, \dots, N^2 -1$, are $N \times N$ matrices that generate the $\mathfrak{su}(N)$ Lie algebra.  
At asymptotic distances, the tension only depends on the $N-$ality of the quark representation. This property can be labeled by an integer $k$ modulo $N$ that dictates how the center of $SU(N)$ is realized 
\begin{equation}{\rm D}(e^{i2\pi/N} I_N) = e^{i2\pi k/N} I_\mathscr{D}   \;,
\end{equation} 
where $I_\mathscr{D}$ is the identity matrix in the $\mathscr{D}$-dimensional representation ${\rm D}$. In a $3D$ spacetime, the asymptotic law 
\begin{equation}
\sigma_D /\sigma_F= k(N-k)/(N-1) 
\label{scaling}  
\end{equation} 
is well-established numerically \cite{Lucini:2001nv}. This behavior corresponds to an asymptotic Casimir law where  the quark representation D in Eq. \eqref{scaling} is replaced by the antisymmetric representation with the same $N-$ality. For a given $k$, the latter gives the lowest quadratic Casimir operator. 
Understanding the mechanism behind these nonperturbative results is a challenging problem.   
 
A remarkably successful approach explored in the lattice is to look for relevant configurations in the infrared. This led to the idea that center vortices and chains, where center-vortex lines change the Lie algebra orientation at lower dimensional defects, are the dominant degrees of freedom (d.o.f.) in this regime \cite{3d,greensite-livro,
Deb+97,Reinhardt:2001kf,
reinhardt-engelhardt,Langfeld:1997jx,DelDebbio:1998luz,FGO98,deForcrand:1999our,Ambjorn:1999ym,Engelhardt:1999fd,BerEngFab01,Gattnar:2004gx}. In particular, the $N-$ality properties of the asymptotic Wilson loop are nicely captured by the center-vortex component \cite{tHooft:1977nqb,Cornwall:1979hz,Mack:1978rq,Nielsen:1979xu}. Until now, a picture for Casimir scaling was only given at intermediate distances \cite{FGO98}, however these arguments depend on the finite size of the center vortex defects, so they cannot be extended to the asymptotic region. In order to incorporate this non-Abelian feature, it is natural to equip the ensemble with non-Abelian d.o.f. Indeed, in the continuum, the presence of these degrees in Yang-Mills theories was brought up in Ref. \cite{oxmanrecent}, noting that the definition of a path-integral measure that detects magnetic defects contains, for every realization of their locations, a continuum of physically inequivalent sectors. In that reference, a class of $4$D Yang-Mills-Higgs effective models was also associated with a $4D$ ensemble of percolating center-vortex worldsurfaces and monopole worldlines with non-Abelian d.o.f. In that case, closed vortices were generated by a dual Yang-Mills term, while the monopoles, carrying adjoint charges, were related to $N^2-1$ flavors of adjoint Higgs fields.  Later, this class of models was shown to be capable of reproducing an asymptotic Casimir law \cite{gustavoxman}.  
   
In this work, we will follow a similar path in $3D$. Initially, we shall propose an ensemble measure where center-vortex worldlines equipped with non-Abelian d.o.f. can be attached to pointlike defects (instantons).  
 This will be an extension of the center vortex measure in Ref. \cite{Oxman-Reinhardt-2017} (see also Refs. \cite{deLemos:2011ww,GBO}), where a $3D$ ensemble of Abelian loops was considered. In that case, in the percolating phase, the effective theory is equivalent to an XY model with topological frustration, which implies an area law for the center-element average. For the extended measure, we will derive an effective field description where the vortices, carrying fundamental weights, get naturally represented by $N$ flavors of effective Higgs fields transforming in the fundamental representation. Up to this point, in the center-vortex condensate, the vacuum manifold has an $SU(N)$ degeneracy. However, the instanton sector is manifested as an additional interaction, which replaces this continuum of possibilities by a discrete set of $Z(N)$ vacua. This gives rise to the formation of a stable domain wall whose border is given by the quark loop. Thus, besides a linear term, the potential will contain a subleading L\"uscher term originated, as usual,  from the fluctuation of collective coordinates around the saddle-point. Finally, after a detailed analysis of the associated field equations, we will show that the string tension turns out to scale with the sought-after asymptotic Casimir law.

 \section{Center-vortices  with non-Abelian d.o.f. }   
 \label{cv-dof}

In Ref. \cite{tHooft:1977nqb}, an Abelian model to describe center vortices in $(2+1)$D was proposed. For this aim, center-vortex operators $\hat{V}(x)$ were defined in $SU(N)$ pure YM theory. In this case, the nontrivial correlators are not only of the form 
    \[
    \langle \Omega | T \{ \hat{V}^\dagger \hat{V} \} | \Omega \rangle 
    \makebox[.5in]{,}  \langle \Omega | T \{ \hat{V}^\dagger \hat{V} \hat{V}^\dagger \hat{V} \} | \Omega \rangle \makebox[.5in]{,} \dots
    \]  
    but also those involving $N$ elementary operators of the same type,
    \[
    \langle \Omega | T \{  \hat{V} \dots \hat{V}\} | \Omega \rangle \makebox[.5in]{,} 
    \langle \Omega | T \{  \hat{V}^\dagger \dots \hat{V}^\dagger\} | \Omega \rangle  \;.
    \]
    This is due to the fact that $N$ center-vortex operators have trivial total $Z(N)$ charge, so they can connect vacuum to vacuum amplitudes. Based on these physical inputs, 
    the following effective model was  then  introduced  
\begin{align}
    {\cal L} =\partial^\mu \bar{V}\, \partial_\mu V + m^2\, \bar{V} V + \frac{\lambda}{2} \, (\bar{V} V)^2 + \xi \, (V^N+\bar{V}^N)  \;, \label{thooft-mod} 
\end{align} 
which captures the above mentioned correlators.
When a condensate is formed ($m^2 < 0$),  the $Z(N)$ symmetry is spontaneously broken, thus leading to classical topological solutions (one-dimensional domain walls) on the physical  
$\mathbb{R}^2$-plane,  with finite energy per unit length. As discussed in Ref. \cite{tHooft:1977nqb}, these line defects can end at a pair of heavy quark-antiquark probes. That is, a confining string is formed in this phase.  

In Ref. \cite{deLemos:2011ww},  the application of polymer techniques to vortex loops with stiffness $1/\kappa$ and tension $\mu$, coupled to an external vector field, made it possible to think of the end-to-end probability of a vortex worldline as a solution to a diffusion equation in $3$D. In this manner, the $N=2$ model in Eq. \eqref{thooft-mod} was associated with the large distance effective description of an ensemble where vortex pairs are created/annihilated via pointlike correlated defects.
In Ref. \cite{Oxman-Reinhardt-2017}, an ensemble measure to compute center-element averages was clearly related with the first three $U(1)$-symmetric terms in Eq. \eqref{thooft-mod},  with a covariant derivative in the place of $\partial_\mu$. This derivative depends on the external field used to represent linking numbers between center vortices and the Wilson loop in the initial ensemble. The  $U(1)$-symmetric sector is dominated by Goldstone modes in a 3d XY model with topological frustration. For large $N$, an analysis based on the associated critical properties led to a {\it squared} sine (area) law. Here, the inclusion of $N$-line correlations is expected to reproduce the complete model in Eq. \eqref{thooft-mod}; however, this cannot accommodate the asymptotic Casimir law either.

In order to describe this type of scaling, which involves a non-Abelian property, it is natural to improve the ensemble with non-Abelian information. Indeed, as discussed in Ref. \cite{OS-det}, the inequivalent sectors of magnetic defects in Yang-Mills theories are naturally labeled by these degrees.  
Therefore, we shall initially  consider the center-vortex loop ensemble of Ref. \cite{Oxman-Reinhardt-2017} embedded in a non-Abelian setting. The  Wilson Loop associated to a quark worldline $\mathcal{C}$ carrying an irreducible $\mathscr{D}$-dimensional representation $D$ is given by
\begin{eqnarray} 
{W}_{\mathcal{C}}[A_\mu]  =  \frac{1}{\mathscr{D}} \, {\rm tr}\, {\rm D}  \left( P \left\{ e^{i \oint_{\mathcal{C}} dx_\mu\,  A_{\mu}(x)  } \right\} \right) \;.
\label{Wloop}   
\end{eqnarray}
The contribution   of an elementary center vortex configuration $A_\mu=\beta \cdot T\, \partial_\mu \chi $, whose field strength is localized on a loop $l$, is the center element
\begin{equation}
W_{\mathcal{C}} [A_\mu]=     
 \left(e^{i \, 2\pi\, \beta \cdot  w_{\rm e} }\right)^{L({\mathcal C}, l) } =
  \left(e^{i2\pi k/N}   \right)^{L({\mathcal C}, l) } 
 \makebox[.5in]{,}  \beta = 2N  w \;,   
 \label{cent-elem}
\end{equation}   
where $L({\mathcal C}, l)$ is the linking number between $\mathcal{C}$ and $l$. The tuple $ \beta$ is a (magnetic) weight of the defining representation, corresponding to unit-flux vortices (there are $N$ possibilities $ \beta_i = 2N  w_i$), and $ \chi$ is a multivalued angle that changes by $2\pi$ when we go around  $l$. These weights can be ordered as $\omega_1>\omega_2>...>\omega_N$.  They satisfy
\begin{equation}
\label{weightproduct}
    \omega_q\cdot \omega_p = \frac{N \delta_{qp}-1}{2N^2}\;.
\end{equation} 
In addition, $ w_{\rm e}$ is an electric weight of the quark representation D.\footnote{For a general representation, a weight $ \lambda$ is an $(N-1)$-tuple formed by the eigenvalues $\lambda|_q$ of simultaneous eigenvectors of the Cartan generators 
${\rm D} (T_q)$, that is, ${\rm D}(T_q) | \lambda \rangle =  \lambda|_q  | \lambda \rangle $, $q=1, \dots, N-1 $. We also defined $ \beta \cdot  T =  \beta|_q  T_q $.} 
This contribution can be rewritten in the form 
\begin{equation}
W_{\mathcal{C}} [A_\mu] = W_{l}[b^{\mathcal C}_\mu] =\frac{1}{N} {\rm tr} P\left( e^{i\oint_l dx_\mu \,  b^{\mathcal C}_\mu(x)}\right) \makebox[.5in]{,} A_\mu=\beta \cdot T\, \partial_\mu \chi  \;,
\label{WAb}
\end{equation}
where $b^{\mathcal C}_\mu(x)\equiv  2\pi \beta_e \cdot T\, s_\mu$, $\beta_{\rm e} = 2N w_{\rm e} $,
 and
 $s_\mu$ is concentrated on a surface $S(\mathcal{C})$ whose border is $\mathcal{C}$ 
\begin{align}
    s_\mu = \frac{1}{2}\int_{S(C)}d\sigma_1d\sigma_2\,\epsilon_{\mu\nu\rho}\frac{\partial x^\nu}{\partial\sigma_1}\frac{\partial x^\rho}{\partial\sigma_2}\delta(\bar{x}(\sigma_1,\sigma_2)-x)\;.
\end{align}
In this respect, note that the circulation of $s_\mu$ along $l$ gives  the intersection number between $l$ and $S(\mathcal{C})$,
\begin{align}
    \oint_l dx_\mu s_\mu = I(S(\mathcal{C}),l) \;,
\end{align}
which coincides with the linking number $L({\mathcal C}, l)$.

 The contribution to the Wilson Loop originated from $n$ center vortices is the product of the corresponding center elements.  
Then, including  the property of stiffness, observed in the lattice \cite{random-surf-1,random-surf-2,random-surf-3}, as well as tension $\mu$, the ensemble average becomes $Z_{\rm loops}[b^{\mathcal C}_\mu]$, where
\begin{eqnarray}
 Z_{\rm loops}[b_\mu] =\sum_{n=0}^\infty \frac{1}{n!}\, \prod_{k=1 }^n   \int_{0}^{\infty}\;
\frac{dL_{k}}{L_{k}}  \int\; dv_k  \int  [dx^{(k)}]_{v_k,v_k}^{L_k}  \,  e^{- \int_0^{L_k}  ds_k\,  \left[  \frac{1}{2\kappa}\, \dot{u}^{(k)}_\mu 
\dot{u}^{(k)}_\mu +
 \mu \right] } \, W_{l_k}[b_\mu]  \;.
 \label{ens-ad}
\end{eqnarray}
is the ensemble partition function in the presence of a general external source $b_\mu \in \mathfrak{su}(N)$. 
For each closed worldline $x^{(k)}(s)$, $u^{(k)}_\mu$ is its tangent vector 
\begin{gather}
 u_\mu(s) =\frac{dx_\mu}{ds}  \in S^2  \makebox[.5in]{,}   \dot{u}_\mu(s) =\frac{du_\mu}{ds}  \;.
\end{gather}  
The loops start and end at some point $x_k \in \mathbb{R}^3$ where the tangent is $u_k \in S^2$. $[dx]_{v,v}^{L}$ path-integrates over loops with length $L$ starting and ending at $v$, $v= (x, u)$.  This partition function
may be rewritten as
\begin{gather} 
Z_{\rm loops} [b_\mu]= e^{  \int_{0}^{\infty}\frac{dL}{L}  \int  dv \,  {\rm tr}\, Q(v, v,L)  }  \;,
\label{pfun}  \\
 Q(x,u,x_0,u_0,L)  =  \int  [dx(s)]_{v,v_0}^L \, e^{- \int_0^{L}  ds\,  \left[  \frac{1}{2\kappa}\, \dot{u}_\mu 
\dot{u}_\mu +
 \mu \right] } \, \Gamma_\gamma[b_\mu] 
 \label{mat-con}  \;, \\
\Gamma_\gamma[b_\mu]   =  P \left\{ e^{i\int_\gamma dx_\mu \,  b_\mu(x)  } \right\} \;.
\end{gather}
We can  use the methods of Refs. \cite{GBO,ref24} to obtain the non-Abelian difusion equation
\begin{align}
&\left(\partial_L-\frac{\kappa}{2}\hat{L}^2_u+\mu+u_\mu(\partial_\mu-ib_\mu)\right)Q(x,u,x_0,u_0,L)=0 \, ,\label{diffu}
\end{align}
to be solved with the initial condition 
$Q(x,u,x_0,u_0,0)= \delta^{(3)}(x-x_0) \delta^{(2)}(u-u_0) I_N$. In the small stiffness limit (large $\kappa$), the solution can be approximated by 
\begin{align} 
Q(x,u,x_0,u_0,L) \approx \langle x | e^{-LO} | x_0 \rangle \makebox[.5in]{,} 
 O = -\frac{1}{3\kappa}(\partial_\mu-ib_\mu)^2+\mu I_N  \;, \label{green} 
\end{align}  
thus leading to
\begin{align} 
Z_{\rm loops}[b_\mu]\approx e^{-{\rm Tr} \, \ln O }= (\det \, O)^{-1} =\int [d\phi]\, e^{-\int d^3x\, \phi^\dagger O \phi}\;, \label{zloops}
\end{align}
where $\phi$ is a complex field in the fundamental representation.   

\subsection{$N$-worldline matching rules}  
 
\begin{figure} 
\centering
\includegraphics[scale=0.45]{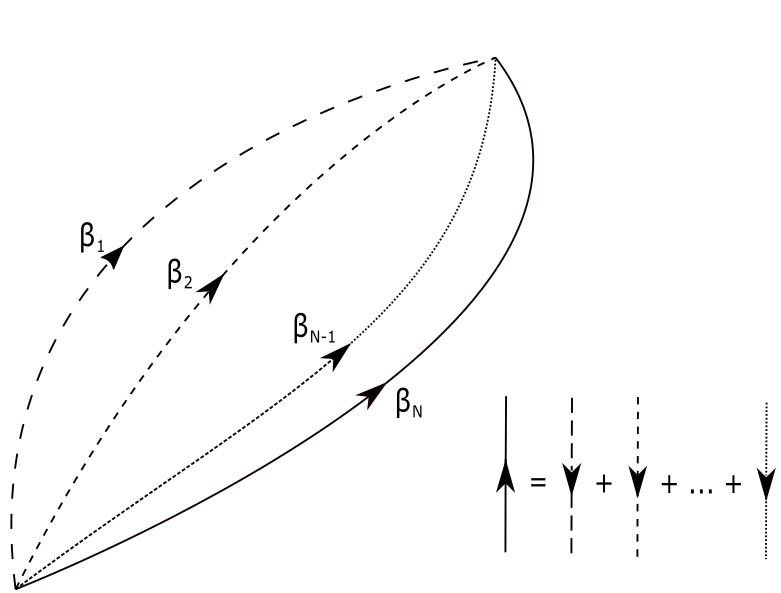}
\vspace{.3cm}
\caption{An $N$ center-vortex creation-annihilation process can be thought of as $N-1$ correlated loops. The loop $l_i$ is formed by joining $\gamma_i$ with $-\gamma_N$.}
\label{corr} 
\end{figure}

The ensemble above can be further improved by including matching rules where $N$ elementary center vortices carrying different weights $\beta_i$, $\sum_i^N \beta_i =0$, can be created at a point out of the vacuum, then propagated along the lines $\gamma_1, \dots, \gamma_N$, and finally annihilated. Since the weights sum to zero, we can think of this configuration as $N-1$ correlated loops $
l_i = \gamma_i -\gamma_N$ with a common line $-\gamma_N$ (see Fig. \ref{corr}), carrying magnetic weights $\beta_i$, $i=1, \dots, N-1$, respectively.   
For this configuration, the gauge field  can be written as $A_\mu= \sum_{i=1}^{N-1} \beta_i \cdot T\, \partial_\mu \chi_{i} $, where $ \chi_{i }$ changes by $2\pi$ when we go around  $l_{i }$. The corresponding contribution to  the Wilson Loop is 
\begin{eqnarray}
W_{\mathcal{C}}[A_\mu] = \big( e^{i \, 2\pi\,  \beta_1 \cdot  w_{\rm e} } \big)^{L(S({\cal C}), l_{1}) }\dots\big( e^{i \, 2\pi\,  \beta_{N-1} \cdot  w_{\rm e} } \big)^{L(S({\cal C}), l_{(N-1)}) }  \;.
\label{wlines}
\end{eqnarray}
 This is a product of center elements that can be rewritten as 
\begin{eqnarray} 
W_{\mathcal{C}}[A_\mu] = \frac{1}{N!}\, \epsilon_{i_1  \dots i_N}\epsilon_{i'_1  \dots i'_N}
\Gamma_{\gamma_1}[b_\mu^{\mathcal C}] |_{i_1 i'_1}\dots\Gamma_{\gamma_N}[b_\mu^{\mathcal C}] |_{i_N i'_N} \makebox[.5in]{,}  A_\mu= \sum_{i=1}^{N-1} \beta_i \cdot T\, \partial_\mu \chi_{i} \;. 
\label{eq1}
\end{eqnarray}
In this regard, since $\langle w_{i} |   \beta_e . T |  w_j \rangle =   w_{\rm e} \cdot \beta_i\, \delta_{ij}$, we have  $e^{2\pi i\int_{\gamma} dx_\mu w_{\rm e} \cdot T\, s_\mu}|_{i j}=e^{2\pi i\int_{\gamma} dx_\mu w_{\rm e} \cdot \beta_i\, s_\mu}\delta_{i j}$, so that the right-hand side in Eq. \eqref{eq1}  becomes,
 \begin{align}
   \frac{1}{N!}\, \epsilon_{i_1  \dots i_N}\epsilon_{i'_1  \dots i'_N}   e^{2\pi i\int_{\gamma_1} dx_\mu w_{\rm e} \cdot \beta_{i_1}\, s_\mu}\delta_{i_1 i'_1}\dots e^{2\pi i\int_{\gamma_N} dx_\mu w_{\rm e} \cdot \beta_{i_N}\, s_\mu}\delta_{i_N i'_N}  \nonumber \\
=      e^{2\pi i\int_{\gamma_1} dx_\mu w_{\rm e} \cdot \beta_1\, s_\mu}\dots e^{2\pi i\int_{\gamma_N} dx_\mu w_{\rm e} \cdot \beta_N\, s_\mu} \nonumber \\
=      e^{2\pi i\oint_{l_1} dx_\mu w_{\rm e} \cdot \beta_1\, s_\mu}\dots e^{2\pi i\oint_{l_{N-1}} dx_\mu w_{\rm e} \cdot \beta_{N-1}\, s_\mu} \;,  
 \end{align} 
which coincides with the right-hand side in Eq. \eqref{wlines}. Including the phenomenological properties of center vortices, the $N$-line configuration gives a contribution:  
\begin{eqnarray} 
&& C_N \propto  \int d^4x \, d^4x_0    \prod_{j=1 }^N  \int dL_{j} du^j  du_0^j  \int [Dx^{(j)}]_{v_0^j \, v^j}^{L_j}  \,  e^{- \int_0^{L_j}  ds_j\,  \left[  \frac{1}{2\kappa}\, \dot{u}^{(j)}_\mu
\dot{u}^{(j)}_\mu + 
 \mu \right] } \, D_N \;, \label{cN} \\
 && D_N =  \epsilon_{i_1\dots i_N}  \epsilon_{j_1\dots j_N} \Gamma_{\gamma_1}[b_\mu]_{i_1 j_1} \dots  \Gamma_{\gamma_N}[b_\mu]_{i_N j_N}  \;. 
 \label{DN}
\end{eqnarray} 
To proceed, using Eqs. \eqref{mat-con} and \eqref{green}, for every line we can use
\begin{equation}
 \int dL\, du \, du_0  \int [Dx]_{v_0 \, v}^{L}  \,  e^{- \int_0^{L}  ds\,  \left[  \frac{1}{2\kappa}\, \dot{u}_\mu 
\dot{u}_\mu +
 \mu\,\right] } \,     \Gamma[b_\mu] =  \int_0^\infty dL \, du\, du_0 \, Q(x , u ,x_0 , u_0,L) \sim  G( x,  x_0 )   \;,
 \label{defgreen}
\end{equation} 
where $
 O\, G(x,x_0) = \delta(x-x_0  ) \, I_{\mathscr{N}} $. In other words,
\begin{equation}
 C_N \propto  \int d^4x \, d^4x_0\,  \epsilon_{i_1\dots i_N}  \epsilon_{j_1\dots j_N} G(x,x_0)_{i_1 j_1}\dots G(x,x_0)_{i_N j_N} \;.
\end{equation}  
The field representation \eqref{zloops} and the discussion above clearly suggest the consideration of $N$ flavors of fundamental fields, one for each fundamental weight,  and an appropriate interaction to accomodate the possible $N$-line matchings by means of the
generated Feynman diagrams in an effective field theory. Indeed, all possibilities for this type of corelation can be generated from the following field partition function
\begin{eqnarray} 
& \int [D\Phi^\dagger] [D\Phi]\, e^{-\int d^3 x\, \left( \frac{1}{3\kappa}{\rm Tr}((D_\mu \Phi)^\dagger D_\mu \Phi) + \mu \,{\rm Tr}( \Phi^\dagger \Phi)  -\xi_0 ( \det \Phi+ \det \Phi^\dagger) \right) } \;,
\label{Z-det}
\end{eqnarray}
where $D_\mu\Phi \equiv (\partial_\mu - ib_\mu) \Phi$ and we defined a matrix with components $\Phi_{i}^{\, j} = \phi^j |_i$, where $\phi^j$, $j=1, \dots, N$ are complex fields in the fundamental representation. A perturbative expansion reads
\begin{eqnarray}
&   \int [D\Phi^\dagger] [D\Phi]\, \big(1+\xi_0^2\int d^3 x\int d^3 x_0\,  
\nonumber\\
&   \epsilon_{i_1  \dots i_N}\epsilon_{i'_1  \dots i'_N} \, \phi^{i'_1}|_{i_1 } \dots \phi^{i'_N}|_{i_N}  \epsilon_{j_1  \dots j_N}\epsilon_{j'_1  \dots j'_N} \, \bar{\phi}^{j'_1}|_{j_1} \dots \bar{\phi}^{j'_N}|_{j_N}+ \dots \big) \,e^{-\int d^3 x\,\bar{\phi}^j|_i O_{ii'}^{jj'}\phi^{j'}|_{i'}} \;.
\end{eqnarray}
Clearly, the first term is  $Z_{\rm loops}^ N$, the contribution to the ensemble  of the uncorrelated $N$ copies of  loop types. 
In addition, multiplying and dividing the $\xi_0^2$-term by
\[
(\det O)^{-N}=\int [D\Phi^\dagger] [D\Phi]\, e^{-\int d^3 x\,\bar{\phi}^j|_i O_{ii'}^{jj'}\phi^{j'}|_{i'}} 
\makebox[.5in]{,}
O^{jj'}_{ii'}\equiv\delta^{jj'}O_{i'i} \;,
\]
and using Wick's theorem, we get 
\begin{eqnarray}
& \frac{\xi_0^2}{(N!)^2}\int [D\Phi^\dagger] [D\Phi]\,\int d^3 x\int d^3 x_0 \, \epsilon_{i_1  \dots i_N}\epsilon_{i'_1  \dots i'_N} \phi^{i'_1}|_{i_1}(x) \dots \phi^{i'_N}|_{i_N}(x)\nonumber \\ 
& \times  \epsilon_{j_1  \dots j_N}\epsilon_{j'_1  \dots j'_N}  \bar{\phi}^{j'_1}|_{j_1}(x_0) \dots\bar{\phi}^{j'_N}|_{j_N}(x_0)e^{-\int d^3 x\,\bar{\phi}^j|_i O_{ii'}^{jj'}\phi^{j'}|_{i'}} \nonumber \\
 & = Z_{\rm loops}^N  \xi_0^2 \int d^3 x\int d^3 x_0 \, \epsilon_{i_1  \dots i_N}  \epsilon_{j_1  \dots j_N}G_{i_1 j_1}(x,x_0)\dots G_{i_N j_N}(x,x_0) \;. 
\end{eqnarray}
Therefore, due to Eqs. \eqref{zloops} and \eqref{defgreen}, we can write this partial contribution as 
\begin{eqnarray}
\xi_0^2   \prod_{j=1 }^N \int d^4x \, d^4x_0 \, \epsilon_{i_1  \dots i_N}  \epsilon_{k_1  \dots k_N}   \int dL_{j} du^j  du_0^j  \int [Dx^{(j)}]_{v_0^j \, v^j}^{L_j}  \,  e^{- \int_0^{L_j}  ds_j\,  \left[  \frac{1}{2\kappa}\, \dot{u}^{(j)}_\mu
\dot{u}^{(j)}_\mu + 
 \mu  \right] }\Gamma^{(j)}_{i_j k_j}[b_\mu]\nonumber\\ \left(\sum_n \frac{1}{n!}\, \prod_{k=1 }^n   \int_{0}^{\infty}\;
\frac{dL_{k}}{L_{k}}  \int\; dv_k  \int  [dx^{(k)}]_{v_k,v_k}^{L_k}  \,  e^{- \int_0^{L_k}  ds_k\,  \left[  \frac{1}{2\kappa}\, \dot{u}^{(k)}_\mu 
\dot{u}^{(k)}_\mu +
 \mu \right] } \, W_{l_k}[b_\mu]\right)^N \;.
\end{eqnarray}
This represents the mixing of the uncorrelated loop configurations and a single correlated two-point component  (cf. Fig. \ref{corr}). Proceeding similarly with the other terms, the perturbative series can be identified with all possible configurations with $N$-line correlations.  
 At this point, the effective model in Eq. \eqref{Z-det} is invariant under (magnetic) local color and global flavor  transformations
\begin{equation}
    \Phi\rightarrow S_c(x)\Phi \makebox[.5in]{,}  b_\mu \to S_c b_\mu S_c^{-1} +i S_c \partial_\mu 
    S_c^{-1}   \makebox[.5in]{,}    \Phi\rightarrow \Phi S_f  \;. 
    \label{c-f}
\end{equation}
Other possible correlations among center vortices, with the same symmetry, can be introduced by means of new terms in the exponent of Eq. \eqref{Z-det}. For example we may consider the center-element average generated by  $Z_{\rm v}[b^{\mathcal C}_\mu]$ where 
\begin{gather}
Z_{\rm v}[b_\mu] = \int [D\Phi^\dagger] [D\Phi]\, e^{-\int d^3 x\,  {\mathcal L}_{\rm v} 
 } \;, \\ 
  {\mathcal L}_{\rm v}  =   \frac{1}{3\kappa}{\rm Tr}((D_\mu \Phi)^\dagger D_\mu \Phi) + \mu \,{\rm Tr}( \Phi^\dagger \Phi) +\lambda_0 {\rm Tr}(\Phi^\dagger \Phi)^2 -\xi_0 ( \det \Phi+ \det \Phi^\dagger) \;,
\label{ele-v}
\end{gather} 
also contains quartic correlations whose importance is weighted by $\lambda $ ($> 0$).

\section{Center vortices with attached instantons}  
  \label{corr-mon} 
  
The non-Abelian model in Eq. \eqref{ele-v} has some points of contact with the Abelian one in Eq. \eqref{thooft-mod}. In particular, the interaction terms coincide for Abelian-like field configurations of the form $\Phi  \approx V I_N $. Indeed, the $\det \Phi$ and $V^N$ terms have a similar physical origin. However, some remarks are in order: i) the continuum symmetries of $ {\mathcal L}_{\rm v} $ would imply a spontaneous symmetry breaking (SSB) phase with a continuum of vacua, rather than the discrete set in Eq. \eqref{thooft-mod} required to support stable domain walls between quarks (confining strings), ii) the path-integral calculation would involve large fluctuations associated with Goldstone modes, iii) the source $b^{\mathcal C}_\mu$, which is along the Cartan sector, prefers field configurations $\Phi$ with a phase along this sector too. In what follows, the situation stated in the first two points will be modified after introducing the possibility of correlated instantons on top of center vortices. In section \ref{Domain}, this procedure, implemented in a
non-Abelian setting, will also accommodate the third point in a natural way. 

In the lattice, most center vortices contain defects, thus forming chains or nonoriented center vortices \cite{Reinhardt:2001kf}. It is therefore reasonable to expect that they might play an important role for describing all the properties of confinement in a satisfactory way. Similarly to the center-vortex configuration  $A_\mu$ in Eq. \eqref{WAb} (resp. \eqref{eq1}), which can be  
written locally (but not globally) as a pure gauge using the singular phase $S =  e^{i\chi\, 
\beta \cdot T}$ (resp. $e^{i \sum_{i=1}^{N-1} \chi_{i}\, \beta_i \cdot T } $), a chain can also be locally introduced as a transformation with phase
\begin{equation}
 S = e^{i\chi\, \beta \cdot T}\, W(x)   \;.
\label{cenvor}
\end{equation}  
 Because of the multivalued phase, the Cartan factor creates a thin center vortex, while $W(x)$
creates lower dimensional defects (see for example \cite{Reinhardt:2001kf,proceedings,conf-qg,lucho}). In $3$D,  these are pointlike (instanton) defects on the center-vortex worldlines, where the Lie algebra orientation changes. These orientations can be associated with two different fundamental weights $w$, $w'$, 
while $W(x)$ is a different Weyl transformation on each side of the instanton, at the center-vortex branches.

 The properties of each type of defect are reflected in the gauge-invariant dual field strength $f_\mu(A) = \epsilon_{\mu \nu \rho} S^{-1} F_{\nu \rho}(A) S $. Considering a more general case where the mappings $S$  are multiplied on the right by a regular map, $S \to S \tilde{U}^{-1}(x)$,  the field strengths  for a thin center-vortex loop and $N$ matched center-vortex lines in Fig. \ref{corr} are respectively given by 
\begin{gather}
f_\mu(A) = f_\mu(l, g(s), \beta)
  \makebox[.5in]{,} f_\mu(A) =\sum_{i=1}^{N}  f_\mu(\gamma_i , g_i(s_i), \beta_i)   \;.
 \label{fN} 
\end{gather}
where
\begin{gather}
 f_\mu(\gamma, g(s), \beta) =  \int_{\gamma} ds \frac{dx_\mu}{ds} \delta (x-x(s))\, g(s) \beta \cdot T g(s)^{-1}  \makebox[.5in]{,}  g(s) \equiv \tilde{U}(x(s)) \;,
 \label{fdefi} 
\end{gather} 
 In addition, for a chain with a pair of instantons, $f_\mu(A)$ is given by
\begin{gather} 
f_\mu(A) = f_\mu(\gamma, g(s), \beta) +  f_\mu(\gamma', g'(s'), \beta')   \;,
\label{fmus1}
\end{gather}
and, for $N\geq 3$, the three instanton contribution can be written as
\begin{gather} 
 f_\mu(A) = f_\mu(\gamma, g(s), \beta) +  f_\mu(\gamma', g'(s'), \beta') + f_\mu(\gamma'', g''(s''), \beta'')    \;.
\label{fmus2}
\end{gather}

\subsection{Introducing correlated pointlike defects}  

Now, we would like to incorporate in the ensemble the possibility of chains. In section \ref{cv-dof}, to derive the effective model, we used as starting point that the center elements $W_{\mathcal{C}} [A_\mu]$, obtained for a thin center-vortex loop and $N$ matched center-vortex lines, can be respectively cast in the form  (cf. Eqs. \eqref{WAb} and \eqref{eq1}) 
\begin{eqnarray} 
 W_{l}[b^{\mathcal C}_\mu]  \makebox[.7in]{\rm and}   \frac{1}{N!}\, \epsilon_{i_1  \dots i_N}\epsilon_{i'_1  \dots i'_N}
\Gamma_{\gamma_1}[b_\mu^{\mathcal C}] |_{i_1 i'_1}\dots\Gamma_{\gamma_N}[b_\mu^{\mathcal C}] |_{i_N i'_N}  \;.
\label{correl-cv} 
\end{eqnarray}
In order to identify correlators analogous to those in Eq. \eqref{correl-cv}, while keeping in the ensemble the information about the attached instantons, we shall initially consider Gilmore-Perelemov group coherent states $|g,w\rangle = g | w\rangle $ (see Refs. \cite{Klauder,Kondo-coh,KOndo-98}), and the completeness relation to write
\begin{gather}
 W_{l}[b_\mu] = \int d\mu(g) \, \langle g,w| \Gamma_l[b_\mu] |g,w\rangle \;.
\label{int-W}   
\end{gather}
In addition,  the identity \cite{Creutz invariant}  
\begin{gather}
 \int d\mu(g)\, g_{i_1 j_1} \dots g_{i_N j_N} = \frac{1}{N!} \epsilon_{i_1\dots i_N} \epsilon_{j_1\dots j_N}  
\end{gather} 
leads to 
\begin{equation}
 \int d\mu(g)\, |g, w_1 \rangle |_{i_1} \dots|g, w_N \rangle |_{i_N}= \int d\mu(g)\, g_{i_1 j_1} \dots g_{i_N j_N} |w_1\rangle |_{j_1} \dots |w_N \rangle |_{j_N}= \frac{1}{N!} \epsilon_{i_1\dots i_N} \;.
\end{equation}
That is, for every $N$-line factor appearing in the ensemble, like the one in Eq. 
\eqref{DN}, we can use the representation
\begin{gather} 
D_N[b_\mu] =  (N!)^2 \int d\mu(g) d\mu(g_0) \langle g, w_1| \Gamma_{\gamma_1} [b_\mu] |g_0, w_1 \rangle \dots  \langle g, w_N| \Gamma_{\gamma_N}[b_\mu] |g_0, w_N \rangle \;.
\label{int-D}
\end{gather}  
The Eqs. \eqref{int-W} and \eqref{int-D} can be interpreted as associating each loop with a weight and each matched line as corresponding to a different weight. 

\begin{figure} 
\centering 
\includegraphics[scale=.65]{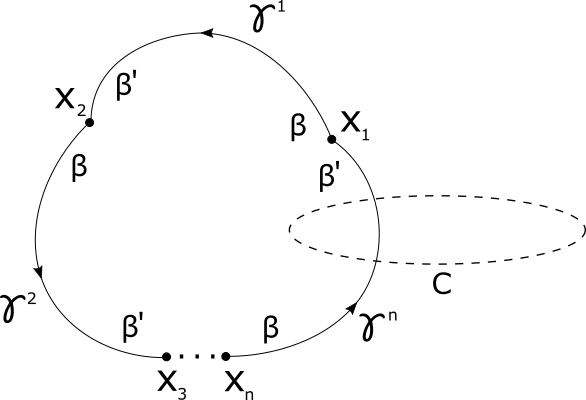}
\vspace{.3cm}
\caption{A chain configuration, with n correlated instantons, linking a Wilson Loop $\mathcal{C}$. }      
\label{chain-2}  
\end{figure}

Then, given that the center-vortex  weight changes at the instantons, to include the effect of chains  with $n$ pointlike defects into the ensemble, we may propose the contribution
\begin{gather}
 \langle g_1 ,w' | \Gamma_{\gamma_n}[b_\mu]|g_{n},w \rangle \dots \langle g_3 ,w' | \Gamma_{\gamma_2}[b_\mu]|g_2,w \rangle \langle g_2 ,w' | \Gamma_{\gamma_1}[b_\mu]|g_1,w \rangle \nonumber \\
= {\rm Tr}\,  \big(  | \Gamma_{\gamma_n}[b_\mu]|g_{n},w \rangle \langle g_n ,w'| \dots  | \Gamma_{\gamma_2}[b_\mu]|g_2,w \rangle \langle g_2 ,w' | \Gamma_{\gamma_1}[b_\mu]|g_1,w \rangle \langle g_1 ,w'| \;
 \label{n-mono1}
\end{gather}
to be integrated over the group elements. However, the integrals of  $g_i|w\rangle\langle w'| g_i^\dagger$ vanish. This follows from the formula
\begin{align}
    \int d\mu(g)\, {\rm D}^{(i)}(g)|_{a b}\,  {\rm D}^{(j)}(g^{-1})|_{  \beta \alpha}    =   \delta_{ij} \delta_{ a \alpha} \delta_{ b \beta}  \;,\label{useful} 
\end{align}
where ${\rm D}^{(i)}$ and ${\rm D}^{(j)}$ are unitary  irreps  \cite{Hammermesh}, applied to the adjoint and trivial irreps.  
 Furthermore, chains also contribute to the Wilson loop  $W_{\mathcal{C}} [A_\mu]$ with a center element \cite{Reinhardt:2001kf,proceedings,conf-qg,lucho,reinhardt-engelhardt}. This comes about as the $W(x)$-factor 
in Eq. \eqref{cenvor} is single-valued when we go around the chain, so that the Wilson loop is only affected by the first factor, which gives the center element in Eq. \eqref{cent-elem}. On the other hand, replacing $b_\mu \to b^{\mathcal C}_\mu$ in Eq. \eqref{n-mono1}, we get a center element times additional overlaps which contain a nontrivial phase. To make sure that the only phases  are associated with center elements, we shall include appropriate overlaps, defining the chain variable (see Fig. \ref{chain-2})
\begin{gather}
\int d\mu(g_1)\dots d\mu(g_n) \, \langle g_1, w|g_2, w' \rangle  \langle g_2, w|g_3, w' \rangle  \dots  \langle g_n, w|g_1, w' \rangle   \nonumber \\
\times \langle g_1 ,w' | \Gamma_{\gamma_n}[b_\mu]|g_{n},w \rangle \dots \langle g_3 ,w' | \Gamma_{\gamma_2}[b_\mu]|g_2,w \rangle \langle g_2 ,w' | \Gamma_{\gamma_1}[b_\mu]|g_1,w \rangle \label{nmon} \\
= \int d\mu(g_1)\dots d\mu(g_n) \, {\rm Tr } \big( |g_n, w' \rangle \langle g_n, w|   \dots |g_2, w' \rangle \langle g_2, w|  \, |g_1, w' \rangle \langle g_1, w|  \big) \nonumber \\
\times {\rm Tr}\,  \big(  | \Gamma_{\gamma_n}[b_\mu]|g_{n},w \rangle \langle g_n ,w'| \dots  | \Gamma_{\gamma_2}[b_\mu]|g_2,w \rangle \langle g_2 ,w' | \Gamma_{\gamma_1}[b_\mu]|g_1,w \rangle \langle g_1 ,w'| \;.
 \label{n-mono2}
\end{gather}
In this manner, for $b_\mu \to b^{\mathcal C}_\mu$, when the chain links $\mathcal{C}$, one of the center-vortex lines will intersect the surface $S(\mathcal{C})$ giving a nontrivial contribution. The final result coincides with the Wilson loop computed for the chain configuration $A_\mu$, times a real and positive weight factor:
\begin{gather}
  \left(e^{i2\pi k/N}   \right)^{L({\mathcal C}, l) }  \int d\mu(g_1)\dots d\mu(g_n) \,  \left| {\rm Tr } \big( |g_n, w' \rangle \langle g_n, w|   \dots |g_2, w' \rangle \langle g_2, w|  \, |g_1, w' \rangle \langle g_1, w|  \big) \right|^2   \;. 
\end{gather}
To obtain an alternative interpretation of the chain and the other defects, we initially note that for, say, the $n=2$ case, we may change $g_2\to g_2 W$, where $W$ is an odd Weyl reflection that takes $w$ into $w'$ and $w'$ and $w$, to get the variable
\begin{gather} 
\int d\mu(g_1)d\mu(g_2) \, \langle g_1, w|g_2, w \rangle  \langle g_2, w'|g_1, w' \rangle  \times   \langle g_1 ,w' | \Gamma_{\gamma_2}[b_\mu]|g_2,w' \rangle \langle g_2 ,w | \Gamma_{\gamma_1}[b_\mu]|g_1,w \rangle \;.\label{2mon}
\end{gather}
Similarly, for $n = 3$, $N>2$ we can make an even Weyl transformation that changes $g_2 \to g_2 P_A$, where $P_A$ permutes $w$, $w'$, $w''$ to $w''$, $w$, $w'$, and then $g_3 \to g_3 P_B$, where $P_B$ permutes $w$, $w'$, $w''$ to $w'$, $w''$, $w$, thus obtaining the variable
\begin{gather}
\int d\mu(g_1) d\mu(g_2)  d\mu(g_3) \, \langle g_1, w|g_2, w \rangle  \langle g_2, w''|g_3, w'' \rangle    \langle g_3, w'|g_1, w' \rangle   \nonumber \\
\times \langle g_1 ,w' | \Gamma_{\gamma_3}[b_\mu]|g_{3},w' \rangle \langle g_3 ,w'' | \Gamma_{\gamma_2}[b_\mu]|g_2,w'' \rangle \langle g_2 ,w | \Gamma_{\gamma_1}[b_\mu]|g_1,w \rangle \;. \label{3mon}  
\end{gather}
Next, we can use the Gilmore-Perelemov representation,    
\begin{align}
\langle g,w |\Gamma_{\gamma}[b_\mu]|g_0,w\rangle=\int [dg(s)]\, e^{i\int ds  {\rm Tr}\left(\left(g(s)^\dagger b(s) g(s) +ig^\dagger(s)  \dot{g}(s)\right)w \cdot T\right)} \makebox[.5in]{,} b(s) 
=   b_\mu(x(s)) \frac{d x_\mu}{ds}  \;,
\label{holo-GP}
\end{align}
where the paths $g(s): [0,L] \to SU(N)$ satisfy the boundary conditions $ g(0)=g_0$, $ g(L)=g $.
 In principle, this applies when the reference state $|w \rangle$ is a highest weight vector. However, the center-vortex holonomy is in the fundamental representation, so the
associated weights can be connected by Weyl transformations. Thus, this formula holds for any weight vector $|w_i \rangle$, $i=1, \dots, N$, which has components $|w_i \rangle|_j = \delta_{ij} $.
As usual (see Refs. \cite{ref26,PD,Perelemov,Gilmore}), the trace in the exponent of Eq. \eqref{holo-GP} can be rewritten in terms of non-Abelian d.o.f. 
$|z(s)\rangle =|g(s), w \rangle$,
\begin{equation}
 {\rm Tr}\left(\dots\right) =
 b^A_\mu(x(s)) T_A|_{ij} \, z_j \bar{z}_i   \frac{dx_\mu}{ds} + \frac{i}{2} (\bar{z}_i \dot{z}_i - \dot{\bar{z}}_i z_i ) \;,
\label{z-var1}
\end{equation}
where the last term can be interpreted as a kinetic term for these degrees. Moreover, using Eq. \eqref{holo-GP} for each line defect, the  $b_\mu$-coupling in the variables containing a loop, $N$ matched center vortex lines, and chains 
with two or three instantons (cf. Eqs. \eqref{int-W}, \eqref{int-D}, \eqref{2mon}, and \eqref{3mon}),
 becomes
   \begin{equation}
e^{i \int d^3x\,  \frac{1}{2N} {\rm Tr}  \big(  b_\mu    f_\mu(A) \big) }      \;,
\label{z-var2}  
\end{equation}
where $f_\mu(A)$ is the gauge-invariant field strength for the corresponding $A_\mu$-configurations  equipped with non-Abelian d.o.f. (see Eqs. \eqref{fN}-\eqref{fmus2}), as described in Ref. \cite{oxmanrecent} when dealing with nonoriented center vortices in $(3+1)$d. That is, the various dual variables, which are designed to reproduce $W[A_\mu]$, precisely couple the corresponding fields $f_\mu(A)$  to $b_\mu(x)$. 

As already discussed, when immersed into the ensemble, the path-integral of the holonomies over paths with tension and stiffness will give rise to a Green's function (cf. Eq. \eqref{defgreen}). In this manner, the chain contributions in Eq. \eqref{nmon} become generated by the new vertex
\begin{align}
V_{\rm inst} \propto  \int d\mu(g) \langle g,w'| \Phi^\dagger |g, w' \rangle 
\langle g,w| \Phi |g, w \rangle,
\end{align}
or, equivalently,
\begin{align}
V_{\rm inst} \propto  
  \int d\mu(g) {\rm Tr}\, \left(  |g, w \rangle \langle g,w'| \Phi^\dagger |g, w' \rangle 
\langle g,w| \Phi \right) \nonumber \\ 
=  \int d\mu(g) {\rm Tr}\, \left(  g| w \rangle \langle w'|g^\dagger \Phi^\dagger g| w' \rangle 
\langle w|g^\dagger \Phi \right) .
\end{align}
Notice that $|w'\rangle \langle w |=E_\alpha$ is a root vector characterized by the root $\alpha=w'-w$. We may write it in terms of the hermitian generators $T_\alpha$, $T_{\bar{\alpha}}$, defined by
\begin{align}
    E_{\alpha}=\frac{T_{\alpha}+iT_{\bar{\alpha}}}{\sqrt{2}}  \;,
\end{align}
and use that $g T_A g^\dagger$ is just the adjoint action of $g$ on $T_A$, $g T_A g^\dagger = R_{A B}(g)T_B $. Then,  
\begin{align}
V_{\rm inst} \propto  \frac{1}{2}\int d\mu(g)   \,{\rm Tr}\left((R_{\alpha B}(g)+iR_{\bar{\alpha} B}(g))T_B\Gamma_\gamma[b_\mu](R_{\alpha C}(g)-iR_{\bar{\alpha} C}(g))T_C\Gamma_{\gamma'}[b_\mu]\right) \;.
\end{align}
To perform the integrals, we can use Eq. \eqref{useful} for the case where $i$ and $j$ stand for the adjoint representation, 
\begin{equation}
\int d\mu(g)\, R_{AB}(g) \, R_{  A'B'}(g)    =   \delta_{ A A'} \delta_{ B B'}  \;.
\label{orth-rel}
\end{equation}
The result is that the instanton-vertex turns out to be
\begin{align}
V_{\rm inst}\propto  {\rm Tr}(\Phi^\dagger T_A\Phi T_A)  \;.
\end{align}

Summarizing, after the discussions in Sec. \ref{cv-dof} and \ref{corr-mon}, we have shown that 
the center element average in the proposed $3$D ensemble, which involves the linking-numbers between the Wilson loop and the mixture of center-vortex loops, correlated $N$-line center vortices and chains, can be effectively represented as  
\begin{equation}
\label{WilsonloopZ}
    \langle W(\mathcal{C})\rangle = \frac{Z[b_\mu^\mathcal{C}]}{Z[0]} 
    \makebox[.5in]{,}
    Z[b_\mu]\equiv \int [D\Phi]\, e^{-S_{\rm eff}(\Phi,b_\mu)}  \;,
\end{equation}    
where the partition function is governed by the large-distance effective action 
\begin{eqnarray} 
\label{action}
    S_{\rm eff}(\Phi,b_\mu) & =& \int\,d^3x \left( {\rm Tr}\, 
    (D_\mu \Phi)^\dagger D^\mu\Phi + V(\Phi, \Phi^\dagger)
     \right) \makebox[.5in]{,}  D_\mu = \partial_\mu -ib_\mu \;, \nonumber \\
&& V(\Phi, \Phi^\dagger) =  \frac{\lambda}{2} {\rm Tr}\,  (\Phi^\dagger\Phi - a^2 I_N)^2  - \xi  \big(\det \Phi + \det\Phi^\dagger\big)-\vartheta    {\rm Tr}\, (\Phi^\dagger T_A \Phi T_A ) + c   \;.  
\end{eqnarray}
Here, we considered a negative tension $\mu$ in Eq. \eqref{ele-v}, which represents a phase where center vortices proliferate. This, together with a positive stiffness $1/\kappa$, implies $a^2 > 0$. This precisely corresponds to a center-vortex condensate. At this point, we notice that the initial color and flavour symmetries of the 
 pure vortex model (c.f. Eq.  \eqref{c-f}) are broken by the instanton term. However, a global color-flavor symmetry ($S_c=S_f^\dagger$) is preserved, as well as a (local) discrete $Z(N)$ symmetry $\Phi \to e^{i\theta_V(x) \beta \cdot T }\Phi$, where $\theta_V(x)$ is a Heaviside function, which is equal to $2\pi$ (resp. $0$) inside (resp. outside) a volume $V$.  The 
latter can be used to change the surface $S({\mathcal C})$ when computing center-element averages. 
The constant $c$ is chosen such that the action at the vacua is zero. In addition, for later convenience, we shall consider a region in parameter space given by positive $\xi $ and $\vartheta  $.

\section{Domain walls with asymptotic Casimir scaling}  
\label{Domain}

In this section, we shall explore the physical consequences of the effective representation for the ensemble of magnetic defects (cf. Eqs. \eqref{WilsonloopZ} and \eqref{action}). For this aim, we shall initially analyze the properties of the spontaneous symmetry breaking phase that the system undergoes. If $N>4$, the potential does not have a lower bound and terms of order higher than $N$ should be included to stabilize it. When seeking a global minimum, we suppose these terms are present, although we do not include them explicitly. We shall consider a region in parameter space so that the potential is dominated by the $\lambda$ and $\xi$-terms.  The polar decomposition $\Phi = PU$, where $P$ is a positive semidefinite hermitian matrix and $U\in U(N) $, can be used to write the potential as
\begin{equation} 
\label{potentialpolar} 
    V (P,U) = \frac{\lambda}{2} {\rm Tr}\, \left( (P^2-a^2I_N)^2 \right) - \xi \det P \, (\det U+\det U^\dagger)-\vartheta   {\rm Tr}\,\left( P T_A P U T_AU^\dagger\right)\;.
\end{equation}
If the only terms were those associated with center-vortex correlations, namely the 
$\lambda$ and $\xi$-terms, then the global minima would certainly be achieved at $P$ proportional to the identity $I_N$. Due to the minus sign, the determinant term also forces $\det U=1$, so that $U\in SU(N)$. Then, up to this point, the possible vacua would form a (continuum) connected manifold, thus precluding the formation of a stable domain wall sitting on the Wilson loop. In this case, the calculation of the partition function would involve large quantum fluctuations associated with the various Goldstone field modes. This type of problem was analyzed in an Abelian context in Ref. \cite{Oxman-Reinhardt-2017}.  On the other hand, when the scenario above is corrected by the inclusion of pointlike defects on center vortices, which is represented by the $\vartheta$-term,  the set of possible vacua becomes discrete.    
   Indeed, because of our choice of sign for $\vartheta$, the minimum values of the potential require a maximum overlap between the basis $T_A$ and the rotated basis $n_A=UT_AU^{-1}$, which is attained when $U$ is in the center $Z(N)$ of $SU(N)$. More precisely, the global minima turn out to be
\begin{subequations}
\begin{align}
P = v I_N\;,\;U\in\mathcal{Z}_N=\left\{\left. e^{i\frac{2\pi n}{N}}I_N\,\right\vert n=0,1,2,...,N-1\right\}\;,\\
2\lambda N(v ^2-a^2) - 2\xi N v ^{N-2}-\vartheta  \left(N^2-1\right) = 0\;. \label{h0} 
\end{align}
\end{subequations}  
Then, it is the presence of correlated instantons that grants the formation of stable domain walls. 
In a $3$D spacetime, a disconnected set of vacua (with nontrivial homotopy group $\Pi_0$) 
enables a field configuration with different vacua on both sides of a surface, with a transition that necessarily implies an action cost localized on the surface. As we will see, in the presence of a Wilson loop, the surface will sit on the loop. Moreover, as the vacua are discrete, there are no Goldstone field modes, and the partition function will be evaluated by means of a saddle point corrected by low-action fluctuations around this point. The former will give rise to a confining area law, while the latter will correct the associated linear potential with the well-known universal L\"uscher term.
A similar situation was recently 
obtained in Ref. \cite{oxmanrecent}, when describing a mixed ensemble of center vortices and chains in $4$D spacetime. In that reference, the inclusion of correlated monopole worldlines on center-vortex worldsurfaces led to a manifold of vacua with nontrivial first homotopy group  $\Pi_1 = Z(N)$. This led to the formation of a confining center string between a quark-antiquark pair, which spans a surface whose border is the Wilson loop. 

Our main objective is to determine the scaling law obeyed by the asymptotic string tension. The saddle-point $\Phi$ for the partition function $Z[b_\mu^\mathcal{C}]$ in Eq. \eqref{WilsonloopZ} satisfies 
\begin{figure}  
\includegraphics[scale=0.35]{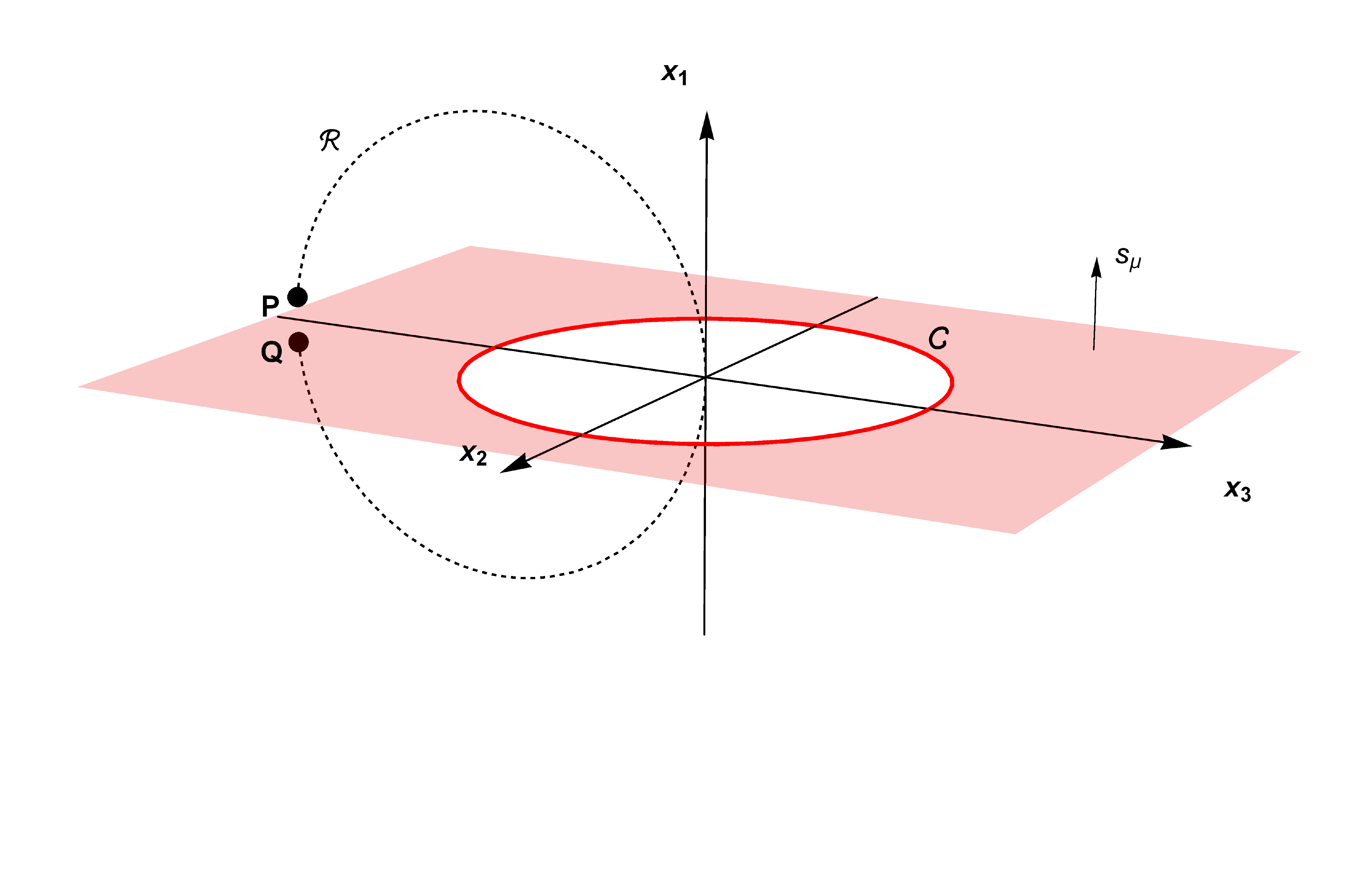}
\vspace{-2.3cm}
\caption{A ring $\mathcal{R}$  winding around the Wilson loop and passing through the origin. The surface where $s_\mu$ is concentrated is depicted in red.}
\label{LargeLoop}
\end{figure} 
\begin{equation} 
\label{Equationphi}
    D^2 \Phi =  \lambda\Phi(\Phi^\dagger\Phi-a^2) - \xi \,C[\Phi^*]-\vartheta   T_A\Phi T_A  \makebox[.5in]{,}  D_\mu = \partial_\mu -i b_\mu^\mathcal{C}\;.
\end{equation}
In this respect, we recall that for small variations of the determinant, we have
\begin{equation}
    \det (\Phi + \delta \Phi) \approx\det\Phi+ {\rm Tr}\, \left(C[\Phi^T]\delta \Phi\right)\;
\end{equation}
where $C[\;]$ stands for the cofactor matrix. Let us analyze how the field $\Phi$ must behave along an arbitrary circle $\mathcal{R}$ that links the loop $\mathcal{C}$. Take for example the loop shown in Fig. \ref{LargeLoop}. From {\bf Q} to {\bf P}, points immediately below and above the intersection between $\mathcal{R}$ and the surface where $s_\mu$ is concentrated, the field $\Phi$ must 'jump' by a phase factor $e^{i2\pi\beta_e\cdot T}$ 
in order to cancel (the regularized form of) $b^C_\mu$ in the covariant derivative, and yield a finite action. This factor is an element of $Z(N)$ and, consequently, the action will be minimized if $\Phi$ is at a vacuum value at {\bf Q}, say $v I_N$, and  continuously changes to the vacuum $v e^{i2\pi\beta_e\cdot T}$ at {\bf P}, as one goes around $\mathcal{R}$. With a discrete set of vacua, this is only possible if $\Phi$ leaves the vacuum somewhere. In general, the transition will be localized around the minimal surface (the disk $\mathbf{D}$) whose border is $\mathcal{C}$. A finite action also requires that above and below the $x_1=0$ plane, and far from the Wilson loop $\mathcal{C}$, the field $\Phi$ must tend to two different vacua, given by the values at {\bf P} and {\bf Q}, respectively. 
In particular, if we follow the $x_1$-axis, or any other parallel line that intersects the disk {\bf D} at coordinates $(0,x_2,x_3)$, the external source leaves 
a trace of its existence only in the boundary condition
\begin{equation}
\label{boundcond}  
   \lim_{x_1 \rightarrow-\infty } \Phi(x_1, x_2, x_3)=v I_N\makebox[.5in]{,}
      \lim_{x_1 \rightarrow +\infty } \Phi(x_1, x_2 , x_3) = v\, e^{i 2\pi \beta_e\cdot T} \makebox[.5in]{,} (0, x_2,x_3) \in \mathbf{D} \;.
\end{equation} 
For an asymptotic Wilson loop, which is much larger than the localization scales in the effective model, the solution will be almost independent of ($x_2 , x_3)$, as long as they remain away from $\mathcal{C}$. In other words, the saddle-point action can be approximated by
\begin{equation}
    S_{\rm eff} \approx  \varepsilon A\;,
\end{equation}
 where $A$ is the area of the disk (plus a border effect that scales as the perimeter), and the string tension is then obtained from the soliton solution $\Phi (x)$,  ($x_1 \equiv x$) that minimizes
 \begin{eqnarray} 
  \varepsilon & =& \int\,dx \left( {\rm Tr}\, 
    (\partial_x \Phi)^\dagger \partial_x\Phi + V(\Phi, \Phi^\dagger)
    \label{ener}
     \right) \;,
\end{eqnarray}
with $\Phi(-\infty)=v I_N$, $\Phi(+\infty) = v\, e^{i 2\pi \beta_e\cdot T}$. This solution satisfies
Eq. \eqref{Equationphi} with the replacement $D^2 \Phi \to \partial_x^2 \Phi $. To close this equation with a simple ansatz, we need to discuss some  properties of the weights of $\mathfrak{su}(N)$.  
For each $N-$ality $k$, we shall consider two types of irreps., called $k-$Symmetric and $k-$Antisymmetric. Their highest weights are $\omega^S_k = k\omega_1$ and $\omega^{A}_k=\omega_1+\omega_2+...+\omega_k$, respectively. Note that, being the sum of $k$ weights of the defining representation, they  yield the correct center element in  Eq. \eqref{cent-elem}.

In the asymptotic regime, gluon screening is expected to take place, bringing down the string tension of any irreducible representation to that of the lowest-dimensional one with the same $N-$ality $k$. The latter corresponds to the $k-$Antisymmetric irrep., $\beta_e = 2N\omega^{A}_k$,  which we shall focus in what follows. In this case, a simple ansatz is motivated by the block-diagonal structure of $\beta_e\cdot T$. If we define $P_1={\rm Diag}(1,1,...,0,0,....0)$ with the first $k$ entries being nonzero and $P_2=I_N-P_1$, we can use Eq. \eqref{weightproduct} to write
\begin{equation} 
\beta_e\cdot T= {\rm Diag} (\beta_e\cdot\omega_1,...,\beta_e\cdot\omega_N) = \left(\begin{matrix}
\frac{N-k}{N}I_k&0 \\
0&-\frac{k}{N}I_{N-k}
\end{matrix}\right) = \frac{N-k}{N} P_1 -\frac{k}{N}P_2\;.
\end{equation}
Because the product between any number of $P_1$ and $P_2$ is either $P_1$, $P_2$ or $0$, an ansatz built upon $P_1$ and $P_2$ will close the equations of motion. Thus, we propose
\begin{equation}
\label{Ansatz}
    \Phi = \left( h_1 P_1 + h_2 P_2\right)  S  \makebox[.5in]{,} S= e^{i\theta_1\frac{N-k}{N}P_1-i\theta_2\frac{k}{N}P_2}\;.
\end{equation}
The phase can be factored in $U(1)$ and $SU(N)$ sectors
\begin{eqnarray}
  S =    e^{i\alpha}e^{i\theta\beta\cdot T} \makebox[.3in]{,} 
    \theta=\frac{N-k}{N}\theta_1+\frac{k}{N}\theta_2 \makebox[.3in]{,}  \alpha = \frac{k(N-k)(\theta_1-\theta_2)}{N^2}\;.
\end{eqnarray} 
In principle, as $e^{i2\pi\beta_e\cdot T}=e^{-i\frac{2k\pi}{N}}$, there are two ways to impose the boundary conditions \eqref{boundcond}: one where $\alpha$ (resp. $\theta$) undergoes a nontrivial transition and leaves the possibility of $\theta$ (resp. $\alpha$)
to remain constant. 
The first possibility gives rise to a model closely related with the `t Hooft's model \cite{tHooft:1977nqb}, for which a Casimir law is not observed, while the second,
\begin{subequations}
\label{boundprofiles} 
\begin{align}
    h_1(-\infty) = h_2(-\infty) = h_0 \makebox[.8in]{,}  h_1(\infty) = h_2(\infty) = h_0 \;,
    \label{boundprofile1}\\
    \theta_1(-\infty) =\theta_2(-\infty)=0 \makebox[.8in]{,}  \theta_1(\infty) = \theta_2(\infty) = 2\pi\;, \\ 
      \theta(-\infty)=0 \makebox[.2in]{,}  \theta(\infty)=2\pi \makebox[.7in]{,} 
    \alpha(-\infty)=0 \makebox[.2in]{,}  \alpha(\infty)=0\;,
    \label{boundprofile2}
\end{align}
\end{subequations}
which is consistent with our choice of external source $b_\mu^\mathcal{C}$, is the option we shall further explore. Moreover, we shall assume $\xi  v^{N-2}>>\vartheta  $, thus disfavoring  $\alpha$  to leave its constant value $\alpha=0$. 
Plugging the ansatz in eq. \eqref{Equationphi} and equating to zero the coefficients of $P_1S$, $P_2S$ $iP_1S$ and $iP_2S$, we obtain
\begin{subequations}
\begin{align}
    \partial_x^2 \, h_1 =& \left(\frac{N-k}{N}\right)^2  (\partial_x\theta_1)^2 h_1 + \lambda  h_1 (h_1^2-a^2) -\xi  h_1^{k-1}h_2^{N-k}\cos\left(\frac{k(N-k)(\theta_1-\theta_2)}{N}\right)\notag\\
    &-\vartheta  \frac{Nk-1}{2N^2}h_1-\vartheta  \frac{N-k}{2N} h_2\cos\left(\frac{k}{N}\theta_2+\frac{N-k}{N}\theta_1\right)\;,\\
     \partial_x^2 \, h_2  =& \left(\frac{k}{N}\right)^2( \partial_x\theta_2)^2 h_2+ \lambda  h_2 (h_2^2-a^2) -\xi h_1^{k}h_2^{N-k-1}\cos\left(\frac{k(N-k)(\theta_1-\theta_2)}{N}\right)\notag\\
    &-\vartheta  \frac{N(N-k)-1}{2N^2}h_2-\vartheta   \frac{k}{2N}h_1\cos\left(\frac{k}{N}\theta_2+\frac{N-k}{N}\theta_1\right)\;,\\
     \partial_x^2 \, \theta_1 =& -2 \partial_x \ln h_1 \,  \partial_x \theta_1 +\xi \frac{N}{N-k}h_1^{k-2}h_2^{N-k} \sin\left(\frac{k(N-k)(\theta_1-\theta_2)}{N}\right)\notag\\
    &+\frac{\vartheta  }{2} \frac{h_2}{h_1}\sin\left(\frac{k}{N}\theta_2+\frac{N-k}{N}\theta_1\right)\;,\\
  \partial_x^2 \,\theta_2 =& -2 \partial_x \ln h_2   \,   \partial_x \theta_2-\xi \frac{N}{k}h_1^kh_2^{N-k-2} \sin\left(\frac{k(N-k)(\theta_1-\theta_2)}{N}\right)\notag\\
    &+\frac{\vartheta  }{2} \frac{h_1}{h_2}\sin\left(\frac{k}{N}\theta_2+\frac{N-k}{N}\theta_1\right)\;.
\end{align}
\end{subequations}
The ansatz in Eq. \eqref{Ansatz} can be rewritten as 
\begin{align}
    \Phi = \left(\eta I_N+\eta_0\beta\cdot T\right)e^{i\theta\beta\cdot T}e^{i\alpha} \makebox[.5in]{,} 
        \eta = \frac{k}{N}h_1+\frac{N-k}{N}h_2 \makebox[.2in]{,}  \eta_0 =  h_1-h_2\;. 
\label{Physicalmodes}
\end{align}  
The equations for these profiles are a little bit more intricate to write down, but they are more meaningful. In particular, if we look at small perturbations around their vacuum value and keep up to linear terms, we get
\begin{subequations}
\label{SmallPerturbation}
\begin{align}
 \partial_x^2\, \delta \eta &= M^2_\eta\,  \delta\eta \makebox[.5in]{,} M^2_\eta = \lambda (3 v ^2-a^2) - \xi  (N-1)v ^{N-2}-\vartheta  \frac{N^2-1}{2N^2} \;, \\
 \partial_x^2\, \delta\eta_0&= M^2_{\eta_0}\,\delta\eta_0\makebox[.5in]{,}M^2_{\eta_0}=\lambda (3v ^2-a^2)+ \xi  v ^{N-2}+\frac{\vartheta  }{2N^2} \;, \\
 \partial_x^2\, \delta\alpha &= M^2_\alpha\,\delta\alpha\makebox[.5in]{,}M^2_\alpha=N \xi  v ^{N-2}\;, \\
 \partial_x^2\,\delta\theta &= M^2_\theta\,\delta\theta\makebox[.5in]{,}M^2_\theta=\frac{\vartheta  }{2}\;. 
\end{align}
\end{subequations}
 These squared masses are non negative (cf. Eq. \eqref{h0}), with  $M_\eta$, $M_{\eta_0}$ and $M_\alpha$  larger than $M_\theta$ due to our previous requirements $\lambda a^2,\,\xi v^{N-2}>>\vartheta$. In this region of parameter space, the functions $\eta$, $\eta_0$ and $\alpha$ are practically constant and $\theta$ is the only one that varies appreciably as it is compelled by the boundary conditions. If the instantons were absent ($\vartheta =0$), the field $\theta$ would be a massless mode associated with the residual $SU(N)$ symmetry of the vacuum. On the other hand, their presence on top of center vortices to form nonoriented center vortices makes the profile $\theta$ to be governed by the Sine-Gordon equation
\begin{equation}
     \partial_x^2\, \theta=\frac{\vartheta  }{2}\sin\theta\;.
\end{equation}
For the soliton solution, we can use Derrick's theorem in Eq. \eqref{ener}  to equate its kinetic and potential contribution so that the string tension in the $k$-Antisymmetric representation is
\begin{equation}
\label{energy} 
 \varepsilon_k = 2 \int dx\left(\left(\frac{N-k}{N}\right)^2 h_1^2 ( \partial_x \theta_1)^2 \,{\rm Tr}\, P_1+ ( \partial_x h_1)^2 \,{\rm Tr}\, P_1 +\left(\frac{k}{N}\right)^2 h_2^2 ( \partial_x \theta_2)^2 \, {\rm Tr}\,P_2+ ( \partial_x h_2)^2 \, {\rm Tr}\,P_2\right)\;,
\end{equation} 
which can be approximated by
\begin{equation}
    \varepsilon_k = \frac{k(N-k)}{N-1}\left(2 v ^2\frac{N-1}{N} \int ( \partial_x \theta)^2 dx\right)  = \frac{k(N-k)}{N-1}\varepsilon_1\;.
\end{equation}
where $\varepsilon_1$ is proportional to the Sine-Gordon parameter $\vartheta$.  
Therefore, the string tension follows a Casimir law. This result can be understood if one considers that, for $\alpha$, $\eta$ and $\eta_0$ frozen at their vacuum value, the only relevant mode is $\theta$ with $\Phi=v e^{i\theta\beta_e\cdot T}$. Consequently, since the total energy is twice the kinetic energy, we get\footnote{Here, we use the normalization ${\rm Tr}\, (T_q T_p)=\frac{\delta_{qp}}{2N}$, which is consistent with Eq. \eqref{weightproduct}.}
\begin{equation}
\label{Energybeta}
    \varepsilon_k = 2v^2 \int {\rm Tr}\big( \partial_x S^\dagger  \partial_x S\big)\, dx = v^2\frac{ \beta_e\cdot\beta_e}{N}\int ( \partial_x\theta)^2 dx\;,
\end{equation}
which, for the $k-$Antisymmetric representation, is proportional to the quadratic Casimir operator:
$\beta_e\cdot\beta_e=2k(N-k)$. For an arbitrary irrep., besides the mode along $\beta_e\cdot T$, additional soft modes in the Cartan sector are needed to close the equations of motion. In this case, a similar procedure can be followed, although it is difficult to analytically obtain the scaling. However, for the $k$-Symmetric representation, it is easy to see that the same ansatz works, and that the energy can be approximated by Eq. \eqref{Energybeta} with 
$\beta_e\cdot\beta_e = 2k^2(N-1)$. This is greater than  
$2k(N-k)$, the value obtained for the  $k-$Antisymmetric case. Therefore, for a given $N$-ality $k$, it
becomes clear that the latter possibility will be preferred, together with its ensuing Casimir law.

\section{Conclusions}

Ensembles of percolating center-vortex  worldlines and worldsurfaces have been detected in Monte Carlo simulations of $SU(N)$ Yang-Mills theory in $3$D and $4$D Euclidean spacetime, respectively. They are relevant degrees that at asymptotic distances provide a Wilson loop area law with $N$-ality. 
A complete picture must also relate this law to the formation of a confining flux tube between a quark and antiquark. Such an object, as well as the effect of its transverse quantum fluctuations, have also been observed. This calls for a field model that supports stable smooth topological objects, with the fields localized around a string in real space. Indeed, relying on different field contents and SSB patterns, many models have been explored in the literature. However, this was mainly done independently of the possible underlying ensembles detected in the simulations.   
In $(3+1)$D, due to Derrick's theorem, besides a vacua manifold for scalars ($\mathcal{M}$) with nontrivial first homotopy group, a gauge field is required to stabilize an infinite\footnote{ This can also be extended to a finite object in the presence of appropriate external quark-antiquark sources.} stringlike soliton in $\mathbb{R}^3$-real space. In this regard, in a recent work \cite{oxmanrecent}, it was satisfying to see  that a $4$D ensemble of percolating center-vortex worldsurfaces with a sector of correlated monopole worldlines can be generated by a dual gauge field with frustration, and a set of adjoint Higgs fields with $\Pi_1(\mathcal{M})= Z(N)$.

 In the present work, we analyzed a similar picture in $(2+1)$D. In $3$D spacetime, center vortices generate worldlines, so they became described by (fundamental) scalars with frustration, rather than by a dual gauge field. In addition, a sector of correlated pointlike defects (instantons) led to a discrete set of vacua ($\Pi_0(\mathcal{M}) = Z(N))$.  Accordingly, in  $(2+1)$D, an infinite (or finite) stringlike soliton in $\mathbb{R}^2$-real space does not require a dynamical gauge field to be stabilized, while the vacua for the scalars must be disconnected.

 Initially, we defined a $3$D measure to compute averages of center elements that depend on the linking number between a Wilson loop and center-vortex worldlines. Modeling these defects with tension and stiffness, we were able to show that, at large distances, center-vortex loops are effectively described by fundamental Higgs fields. On the one hand, this is related to the fact that elementary center vortices carry fundamental weights, on the other, this type of field is originated when taking into account non-Abelian d.o.f. propagated on the worldlines. The possibility of $N$-line center-vortex matching is natural, as the different weights of the fundamental representation add up to zero. This was included by means of $N$ flavors, which can be arranged as an $N\times N$ complex matrix $\Phi$. All possible combinations of loops and correlated lines were generated by an effective theory with (local) $SU(N)$ magnetic color and (global) $SU(N)$ flavor symmetry. The $N$-line matching is responsible for breaking the local $U(N)= U(1)\times SU(N)$ symmetry, that would be present for loops, to $SU(N)$. If this model were restricted to Abelian-like configurations $\Phi = V I_N$, $V \in \mathbb{C}$, we would make contact with the `t Hooft model, where a $U(1)$ symmetry is spontaneously broken to $  Z(N)$, due to presence of a term $V^N + \bar{V}^N$.  However, there is no dynamical basis for such a restriction, and at this point our effective description possesses large quantum fluctuations. Next, we incorporated the effect of chains formed by different center-vortex lines interpolated by pointlike defects. For this aim, the variables used to represent chains were carefully written in terms of dual holonomies, in analogy with the ones describing loops and $N$ matched lines. The immersion of all chain combinations into the ensemble led to an additional effective vertex. In a percolating phase, where large center vortices are favored, a discrete set of vacua was then obtained $\Phi = v  e^{i\frac{2\pi n}{N}}I_N\,,  n=1,2,...,N$, dynamically reducing the $SU(N)$ magnetic color symmetry to the required discrete $Z(N)$. This led to the formation of a stable domain wall sitting on the Wilson loop. Therefore, the center-element average not only displays an asymptotic area law with $N$-ality but it is due to a localized field configuration, which constitutes the interquark confining string.  The potential also contains a subleading universal L\"uscher term associated with the first corrections to the saddle point: the transverse string fluctuations.
The asymptotic string tension for a general antisymmetric representation of $SU(N)$ was then derived by computing the domain wall for a large Wilson loop. The solution is given by a kink that interpolates a pair of different vacua. Furthermore, there is a region in parameter space where the wall is governed by the Cartan sector. In this region, we approximately closed an ansatz to solve the equations of motion and showed that the string tension satisfies the asymptotic Casimir law observed in Monte Carlo simulations of $3$D $SU(N)$ Yang-Mills theory.

Finally, we recall that the  
$4$D YM string transverse field distribution via Monte Carlo \cite{NOprofile}, which is given by a Nielsen-Olesen profile,  can be accommodated  in the effective description of center-vortex worldsurfaces and correlated monopole worldlines \cite{gustavoxman}. It would also be quite interesting to compare the $3$D YM simulation with the distribution obtained in this work, for center-vortex worldlines and correlated instantons, which is close to a Sine-Gordon profile.

\section*{Acknowledgements}

We thank H. Reinhardt, M. Quandt, G. Burgio, and D. Campagnari for useful discussions.
 The Conselho Nacional de Desenvolvimento Cient\'{\i}fico e Tecnol\'{o}gico (CNPq), the Coordena\c c\~ao de Aperfei\c coamento de Pessoal de N\'{\i}vel Superior (CAPES), and the Funda\c c\~{a}o de Amparo \`{a} Pesquisa do Estado do Rio de Janeiro (FAPERJ) are acknowledged for their financial support.


\begin{thebibliography}{99} 
\bibitem{Bali} G. S. Bali, Phys. Rep. {\bf 343} (2001) 1.

\bibitem{Bali2} G. S. Bali, Phys. Rev. {\bf D 62}, (2000) 114503.

\bibitem{Lucini:2001nv} B. Lucini and M. Teper, Phys. Rev. {\bf D64} (2001) 105019.

\bibitem{3d} J. Gattnar, K. Langfeld, A. Schäfke and H. Reinhardt, Phys. Lett. {B489} (2000) 251.
 
\bibitem{greensite-livro} J. Greensite, \textit{An Introduction to the Confinement Problem} ($1^{st}ed.$, Springer, 2011).

\bibitem{Deb+97}
L.~Del~Debbio, M.~Faber, J.~Greensite and S.~Olejnik, Phys. Rev. \textbf{D55} (1997)
  2298.
%\newblock \doi{10.1103/PhysRevD.55.2298}

\bibitem{Reinhardt:2001kf}
H.~Reinhardt, Nucl. Phys. \textbf{B628} (2002) 133.
%\newblock \doi{10.1016/S0550-3213(02)00130-X}

\bibitem{reinhardt-engelhardt} M. Engelhardt and H. Reinhardt, Nucl. Phys. {\bf B567} (2000) 249.

\bibitem{Langfeld:1997jx}
K.~Langfeld, H.~Reinhardt and O.~Tennert, Phys. Lett. \textbf{B419} (1998) 317.
%\newblock \doi{10.1016/S0370-2693(97)01435-4}

\bibitem{DelDebbio:1998luz}
L.~Del~Debbio, M.~Faber, J.~Giedt, J.~Greensite and S.~Olejnik, Phys. Rev.
  \textbf{D58} (1998) 094501.
%\newblock \doi{10.1103/PhysRevD.58.094501}

\bibitem{FGO98} M. Faber, J. Greensite and \v{S}. Olejn\'{\i}k
Phys. Rev. {\textbf D57} (1998) 2603.  

\bibitem{deForcrand:1999our}
P.~de~Forcrand and M.~D'Elia, Phys. Rev. Lett. \textbf{82}  (1999) 4582.
%\newblock \doi{10.1103/PhysRevLett.82.4582}

\bibitem{Ambjorn:1999ym}
J.~Ambjorn, J.~Giedt and J.~Greensite, JHEP \textbf{02} (2000) 033.
%\newblock \doi{10.1088/1126-6708/2000/02/033}

\bibitem{Engelhardt:1999fd}  
M.~Engelhardt, K.~Langfeld, H.~Reinhardt and O.~Tennert, Phys. Rev. \textbf{D61} (2000) 054504.
%\newblock \doi{10.1103/PhysRevD.61.054504} 

\bibitem{BerEngFab01}
R.~Bertle, M.~Engelhardt and M.~Faber, Phys. Rev. \textbf{D64} (2001) 074504.
%\newblock \doi{10.1103/PhysRevD.64.074504} 

\bibitem{Gattnar:2004gx}
J.~Gattnar, C.~Gattringer, K.~Langfeld, H.~Reinhardt, A.~Schafer and S.~Solbrig,
  T.~Tok, Nucl. Phys. \textbf{B716} (2005) 105.
%\newblock \doi{10.1016/j.nuclphysb.2005.03.027}

\bibitem{tHooft:1977nqb}
G.~`t~Hooft, Nucl. Phys. \textbf{B138} (1978) 1.
%%\newblock \doi{10.1016/0550-3213(78)90153-0}

\bibitem{Cornwall:1979hz}
J.M. Cornwall, Nucl. Phys. \textbf{B157} (1979) 392.
%\newblock \doi{10.1016/0550-3213(79)90111-1}

\bibitem{Mack:1978rq}
G.~Mack and V.B. Petkova, Annals Phys. \textbf{123}  (1979) 442.
%\newblock \doi{10.1016/0003-4916(79)90346-4}

\bibitem{Nielsen:1979xu} 
H.B. Nielsen and P.~Olesen, Nucl. Phys. \textbf{B160} (1979) 380.
%\newblock \doi{10.1016/0550-3213(79)90065-8}

\bibitem{oxmanrecent} L. E. Oxman, Phys. Rev {\bf D98} (2018) 036018.

\bibitem{gustavoxman} L. E. Oxman and G. M. Sim\~oes, Phys. Rev. {\bf D99} (2019) 016011.

\bibitem{Oxman-Reinhardt-2017} L. E. Oxman and H. Reinhardt, Eur. Phys. Jour. {\bf C78} (2018) 177.

\bibitem{deLemos:2011ww}
A.L.L. de~Lemos, L.E. Oxman and B.F.I. Teixeira, Phys. Rev. \textbf{D85} (2012) 125014.

\bibitem{GBO} L. E. Oxman, G. C. Santos-Rosa and B. F. I. Teixeira, Jour. Phys. \textbf{A47} (2014) 305401.

\bibitem{OS-det} L. E. Oxman and G. C. Santos-Rosa,
Phys. Rev. {\bf D92} (2015) 125025.

\bibitem{random-surf-1} M. Engelhardt and H. Reinhardt, Nucl. Phys. {\bf B585} (2000) 591.

\bibitem{random-surf-2} M. Engelhardt, M. Quandt and H. Reinhardt, Nucl. Phys. {\bf B685} (2004) 227.

\bibitem{random-surf-3} M. Quandt, H. Reinhardt and M. Engelhardt, Phys. Rev. {\bf D71} (2005) 054026.

\bibitem{ref24} G. H. Fredrickson,  \textit{The Equilibrium Theory of Inhomogeneous Polymers} ($1^{nd}$ed., Clarendon Press, Oxford, 2006), p. 452.

\bibitem{proceedings} H. Reinhardt and M. Engelhardt, \textit{Center vortices in continuum Yang-Mills Theory}, in Proceeding of the 4th International Conference in Quark confinement and the hadron spectrum, Vienna, Austria, July 3-8, 2000.

\bibitem{conf-qg} L. E. Oxman, JHEP \textbf{03} (2013)
038.

\bibitem{lucho} L. E. Oxman, JHEP {\bf 12} (2008) 089.

\bibitem{Klauder} John R. Klauder and Bo-Sture Skagerstam, \textit{Coherent States: Applications in Physics and Mathematical Physics} (World Scientific, 1985).

\bibitem{Kondo-coh} Kei-Ichi Kondo, Phys. Rev. {\bf D58} (1998) 105016.

\bibitem{KOndo-98} Kei-Ichi Kondo and Yutaro Taira, Prog. Theor. Phys. {\bf 104} (2000) 1189.

\bibitem{Creutz invariant} M. Creutz, Journal of Mathematical Physics {\bf 19} (1978) 2043.

\bibitem{Hammermesh} M. Hamermesh, Group Theory and its Applications to Physical Problems (Dover, 1989).


\bibitem{ref26} A. P. Balachandran, P. Salomonson, B. Skagerstam and J. Winnberg, 
 Phys. Rev.  \textbf{D15} (1977) 2308.

\bibitem{PD} D. Diakonov and V. Petrov, Phys. Lett. {\bf B224} (1989) 131.

 \bibitem{Perelemov} A. Perelemov, \textit{Generalized Coherent States and Their Applications} (Springer Verlag, 1986).
 
\bibitem{Gilmore} Wei-Min Zhang, Da Hsuan Feng and Robert Gilmore,
Rev. Mod. Phys. {\bf 62} (1990) 867.
 
\bibitem{NOprofile} P. Cea, L. Cosmai, F. Cuteri, and A. Papa, Phys. Rev.
{\bf D 95} (2017) 114511. 

\end{thebibliography}
\end{document}